\documentclass[a4paper,12pt]{article}

\usepackage[centertags]{amsmath}
\usepackage{amsfonts}
\usepackage{amssymb}
\usepackage{amsthm}
\usepackage{newlfont}

\usepackage[all]{}
\usepackage{graphics}
\usepackage{epsfig}

\newcommand{\lam}{\lambda}
\newcommand{\ort}{\bot}

\newcommand{\eps}{\varepsilon}

\newcommand{\rcl}{R_{\mathrm{cl}}}
\newcommand{\rqu}{\sigma_{\footnotesize{T}}}

\theoremstyle{plain}
  \newtheorem{theorem}{Theorem}[section]

\theoremstyle{definition}
  
  \newtheorem{assumption}[theorem]{Assumption}
\theoremstyle{remark}
  \newtheorem{remark}[theorem]{Remark}

\numberwithin{equation}{section}

\let\si=\sigma

   \let\Om=\Omega

\newcommand{\caI}{{\mathcal I}}

\newcommand{\caO}{{\mathcal O}}

\newcommand{\bbR}{{\mathbb R}}

\newcommand{\opunit}{\text{1}\kern-0.22em\text{l}}

\newcommand{\ra}{\rightarrow}

\newcommand{\e}{{\mathrm e}}
\renewcommand{\i}{{\mathrm i}}
\renewcommand{\d}{{\mathrm d}}

\newcommand{\beq}{ \begin{equation} }
\newcommand{\eeq}{ \end{equation} }
\newcommand{\bet}{ \begin{theorem} }
\newcommand{\eet}{ \end{theorem} }

\newcommand{\refl} {\mathrm{\mathbf{  R }_z }}

            \newcounter{smallarabics}
\newenvironment{arabicenumerate}
{\begin{list}{{\normalfont\textrm{\arabic{smallarabics})}}}
  {\usecounter{smallarabics}\setlength{\itemindent}{0cm}
  \setlength{\leftmargin}{5ex}\setlength{\labelwidth}{4ex}
  \setlength{\topsep}{0.75\parsep}\setlength{\partopsep}{0ex}
   \setlength{\itemsep}{0ex}}}
{\end{list}}

\newcounter{smallroman}

\newcommand{\ben}{\begin{arabicenumerate}}
\newcommand{\een}{\end{arabicenumerate}}

\author{ W.~De Roeck \thanks{FWO- aspirant Universiteit Antwerpen and
K.U.Leuven, Belgium. e-mail : wojciech.deroeck@fys.kuleuven.be}
\qquad E.L.~Lakshtanov \thanks{FCT-posdoc Aveiro University,
Department of Mathematics, Portugal, RAS. e-mail :
lakshtanov@rambler.ru} \quad }

\title{Perturbative estimates on the transport cross section in quantum scattering by hard obstacles}
\date{}

\begin{document}
 \maketitle
\begin{abstract}

The quantum scattering by smooth bodies is considered
for small and large values of $kd$, with $k$ the wavenumber and $d$ the
scale of the body. In both regimes, we prove that the
forward scattering exceeds the backscattering. For high $k$, we need to assume that the body is strictly convex.
\\

\end{abstract}


\section{Introduction}\label{sec: intro}
\subsection{Quantum scattering}
\begin{figure}\label{scat}
\begin{center}
\includegraphics[width=0.60\textwidth, angle=-90]{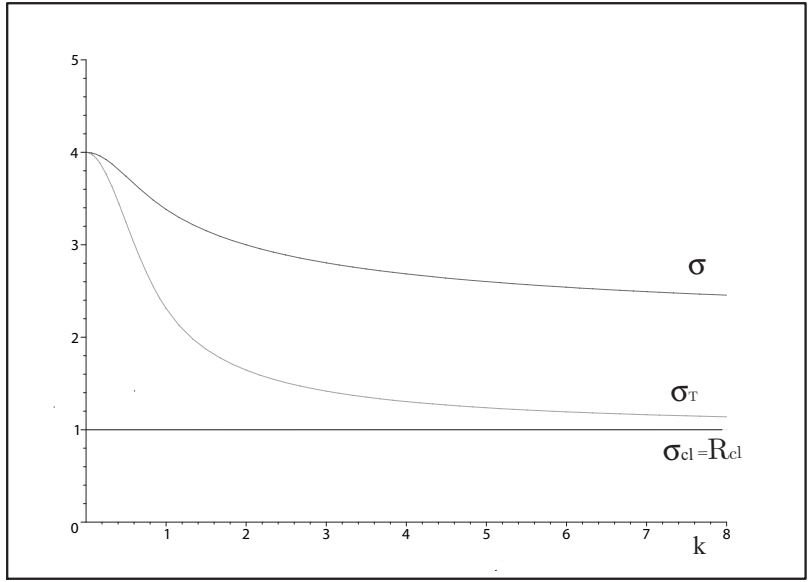}
\parbox[t]{0.47\textwidth}{\caption{Transport cross section $\rqu$ and classical resistance $\rcl$; total cross section $\sigma$ and classical total cross section $\sigma_{\mathrm{cl}}$ for the hard sphere with radius $r= \pi^{-1/2}$.}\label{normRF}}
\end{center}
\end{figure}

We briefly present in physical language the quantum scattering
problem for hard objects in three dimensions. Fix a $z$-axis in
$\bbR^3$ and denote the unit vector along that axis as
$\mathbf{e}=(0,0,1) \in \bbR^3$. Let a body be given as a compact
subset $\Omega \subset \mathbb{R}^3$ and consider a flow of free
quantum particles with wave vector $\mathbf{k}=k \mathbf{e}$,
incident on $\Om$.  The body is modeled by a hardcore potential
$V_{\Omega}$,

\begin{equation}\label{hardspherepot}
V_{\Omega}(\mathbf{r})=\left\{
\begin{array}{ll} 0  & \textrm{ if } \mathbf{r}
\notin \Omega,
\\
+\infty &\textrm{ if } \mathbf{r} \in \Omega .
\end{array} \right.
\end{equation}
Basic scattering theory \cite{Landau} teaches us that far from the scatterer (in
the limit $\mathbf{r} \uparrow \infty$), the wave function
$\Psi(\mathbf{r})$ is obtained by adding an outgoing spherical
wave $\frac{f( \mathbf{q})}{r} \e^{\i kr}$ to the incoming plane
wave $\e^{\i k z}$.
\begin{equation}\label{def: f}
\Psi(\mathbf{r}) \approx \e^{\i k z}+\frac{f( \mathbf{q})}{r}
\e^{\i kr} , \qquad \mathbf{r} \in \bbR^3 \setminus \Omega, \,
\mathbf{q} \in S^2
\end{equation}
where $S^2$ is the unit sphere: $\mathbf{q} \in S^2
\Leftrightarrow \mathbf{q} \cdot \mathbf{q} =1$, $r:=|\mathbf{r}|,
k=|\mathbf{k}|$ and $z = \mathbf{r} \cdot \mathbf{e}$. This
notation will be used throughout the paper. The function $f(
\mathbf{q})$ goes under the name of \emph{scattering amplitude},
it describes the form of the outgoing spherical wave. The
\emph{scattering amplitude} depends on $k d$ where $d$ is the
typical scale of $\Omega$. For simplicity we keep $\Omega$ (and
hence $d$) fixed and we vary $k$. The intensity of the scattered
wave is given by the \emph{total cross section}
\begin{equation}\label{cross}
\sigma=\int_{S^2} \d \mathbf{q} \, |f( \mathbf{q})|^2
\end{equation}
where $\d \mathbf{q}$ is the uniform measure on the sphere. We
also consider the \emph{momentum transfer cross section}, or
\emph{transport cross section} $\sigma_{T}$,
\begin{equation}\label{qRes}
 \sigma_{T}= \left .
\frac{1}{k}\int_{S^2} \d \mathbf{q} \, \mathbf{k}
\cdot(\mathbf{e}-\mathbf{q}) |f( \mathbf{q})|^2  \right.
\end{equation}
Both $\si$ and $\sigma_{T}$ have the dimension of an \emph{area},
justifying the name \emph{cross section}.
They can be computed explicitly for the sphere \cite{sph}, see
Fig. \ref{scat}.
We see that for all positive $k>0$,
\begin{equation} \label{eq: basic inequality}
\sigma_{T}<\sigma
\end{equation}
By some rewriting,
\begin{eqnarray}\label{eqn: difference}
\sigma_{T} -\sigma &=& \int_{S^2} \d \mathbf{q} \, \mathbf{e} \cdot(
\mathbf{e}- \mathbf{q})
|f( \mathbf{q})|^2 -   \int_{S^2} \d \mathbf{q} \, |f( \mathbf{q})|^2 \\
&=& - \int_{S^2} \d \mathbf{q} \, \cos \theta |f( \mathbf{q})|^2
\end{eqnarray}
where $\theta$ denotes the angle between $\mathbf{q}$ and
$\mathbf{e}$. We see that the inequality \eqref{eq: basic inequality}
means that the forward scattering is greater than the
backscattering.

It is a well-known physical fact that at $k=0$, the scattering is isotropic. Indeed, if we write $C(\Om)$ for the capacity of $\Om$ (defined further in
\eqref{def: capa}), then
\begin{equation}
f( \mathbf{q})= -C+ \caO(k)
\end{equation}
This was established rigourously in \cite{Ramm}. An obvious
consequence is that, in lowest order in $k$, the \emph{momentum
transfer cross section} coincides with the \emph{total cross
section},
\begin{equation}
 \sigma_{T}  + \caO(k) =  \si +\caO(k)= 4 \pi C^2.
\end{equation}
Apart from this obvious fact, we know of no place in the
literature where the relation between $\si$ and $\sigma_{T}$ is
examined (quite in contrast to the classical case, see Section
\ref{sec: clas}). More generally, we are not aware of any
qualitative results on the \emph{scattering amplitude} for small
but nonzero $k >0$, other than the optical theorem
 \[
 \frac{4 \pi}{k} \Im f(\mathbf{e}) =  \sigma
 \]
 A natural question seems to be how general the inequality \eqref{eq:
basic inequality} is. Remark that the optical theorem does not
answer this question, although it does say that the forward
scattering cannot vanish completely.

Our first result, Theorem \ref{thm: perturbative low freq},
establishes the inequality $\eqref{eq: basic inequality}$ perturbatively up to
order $k^3$ for a general class of bodies. Our second result,
Theorem  \ref{thm: R < sigma at high k}, establishes the inequality $\eqref{eq:
basic
inequality}$ for large $k$.\\

\subsection{Classical analogue} \label{sec: clas}
We briefly construct the \emph{classical scattering amplitude}
$f_{\mathrm{cl}}$ associated to a body $\Omega$.

 Consider a flow of \emph{classical} particles with
momentum $\mathbf{k}$, incident on $\Om$. The particles will move
freely, then undergo several\footnote{For nonconvex bodies, it can
happen that incoming particles undergo an infinite number of
collisions. Excluding this possibility requires an additional
assumption.} elastic collisions with $\Omega$ and finally move
freely again with momentum $\mathbf{k}^+(\mathbf{x})$ where
$\mathbf{x} \in \mathbb{R}^2$ marks their initial coordinates in
$\mathbf{e}^{\ort}$, the plane perpendicular to $\mathbf{e}$.
Since the collisions are assumed elastic, we have
$|\mathbf{k}^+|=k$.\\
Let $\caI \subset \bbR^2$ be the shadow associated to $\Omega$,
i.e.
\begin{equation}\label{def: shadow}
\mathbf{x} \in \caI \Leftrightarrow   \exists z \in \bbR :
(\mathbf{x},z) \in \Omega
\end{equation}
Let $F$ be the map from  $\mathcal I$ to the sphere $S^2$ such
that $F(\mathbf{x})=\frac{\mathbf{k}^+(\mathbf{x})}{k}$. Assuming
strict convexity of $\Omega$, the inverse $F^{-1}$ exists
(possibly up to a set of measure zero). We define the
\emph{classical scattering amplitude} as \beq f_{\mathrm{cl}}(
\mathbf{q})=|J(F^{-1})( \mathbf{q})|^{1/2}, \qquad \mathbf{q} \in
S^2 \eeq where $J(F^{-1})$ is the Jacobian determinant of the map
$F^{-1}$. Now one can define the \emph{classical resistance}
$R_{\mathrm{cl}}$ and the \emph{classical cross section}
$\si_{\mathrm{cl}}$ in analogy to \eqref{qRes} as
\begin{equation}\label{def: classical R and sigma}
\si_{\mathrm{cl}}= \int_{S^2} \d \mathbf{q} \, | f_{\mathrm{cl}}(
\mathbf{q})|^2 \qquad R_{\mathrm{cl}}= \int_{S^2} \d \mathbf{q} \,
\cos \theta | f_{\mathrm{cl}}( \mathbf{q})|^2
\end{equation}
which is equivalent to the more straightforward definitions
\begin{equation}\label{def: straight}
\sigma_{\mathrm{cl}}=\int_{\caI} \d \mathbf{x} = |\caI | , \qquad
R_{\mathrm{cl}}= \frac{1}{k} \int_{\caI} \d
\mathbf{x} \,\mathbf{e} \cdot
(\mathbf{k}-\mathbf{k}^+(\mathbf{x}))
 \end{equation}
 (In fact, the function $ f_{\mathrm{cl}}$ can be infinite on a set of measure zero, but it remains integrable. This follows e.g.\ by rewriting it as \eqref{def: straight}.)

At this point one can ask some interesting questions: Already
Newton \cite{Newton} posed and solved the problem of minimizing
$R_{\mathrm{cl}}$ in the class of axially symmetric convex bodies
inscribed in a fixed cylinder. Recently, this problem has received
renewed attention, see e.g. \cite{plakhov}. The quantum analogue
of this problem; minimizing $\si_{T} $ while keeping $\si$ fixed,
seems by far out of reach.

\section{Results}\label{sec: results}

Assume for simplicity that $\Om$ is a compact body with smooth surface, i.e.
it is in the class $C^{\infty}$. We rewrite \eqref{def: f} as a
bounday value problem. Let $u$ be
 a function on $\bbR^3 \setminus \Omega$, satisfying
\begin{enumerate} \label{cond: on u}
\item{The Helmholtz equation $(\triangle +k^2 ) u=0 $}
\item{The boundary condition  $ u(\mathbf{x}) = -\e^{\i k z }   $ for $  \mathbf{x} \in \delta \Omega $}
\item{ The Bohr-Sommerfeld radiation criterion \[ \lim_{s \uparrow \infty}\int_{r=s} \d  \mathbf{r} \, (\mathbf{r}\cdot \nabla -\i k  )u=0  \] }
\end{enumerate}
One shows (see e.g.\cite{Ramm}) that these conditions  admit a
unique solution $u$. The \emph{scattering amplitude} $f$ is
defined as \beq \label{connection f and u} f( \mathbf{q}):=
\lim_{r \uparrow +\infty} \e^{-\i k r} r u(r \mathbf{q})  \eeq
We define the capacity $C(\Omega)$ by
\begin{equation} \label{def: capa} C(\Omega) = \int_{\partial \Om} \d \sigma (\mathbf{p}) \, \nu(\mathbf{p}) \quad  \textrm{    with $\nu$
the solution of     } \quad \int_{\partial \Om} \d \sigma
(\mathbf{p}) \, \frac{\nu(\mathbf{p})
}{|\mathbf{p}-\mathbf{r}|}=1, \quad \mathbf{r} \in \partial \Om
\end{equation}
where $\d \sigma(\mathbf{p})$ is the measure on $\partial \Omega$,
inherited from Lesbegue measure on $\bbR^3$.
Our first result, Theorems \ref{thm: perturbative low freq}
speaks about the low frequency regime.
\begin{theorem}\label{thm: perturbative low freq}
Let $\si$ and $\si_T$ be as defined in \eqref{cross} and
\eqref{qRes} with $f(\mathbf{q})$ as defined in \eqref{connection
f and u}. Let $C(\Om)$ be the capacity as in \eqref{def: capa} and
$V(\Om)$ the volume of a smooth compact body $\Om$, then
\begin{equation}\label{eq1}
\si_{T} \leq \sigma- \frac{4 \pi}{3}k^2C(\Omega)V(\Omega) + \caO(k^4)
\end{equation}
\end{theorem}
%
%\begin{remark}
%Remark that the relation between $V$ and $C$ is constrained by the
%Poincar\'e-Faber-Szeg\"o inequality \cite{polya}:
%\begin{equation}
%V(\Omega) \leq \frac{4 \pi}{3} C(\Omega)^3
%\end{equation}
%where equality is reached for the sphere.
%\end{remark}

This follows by application of standard Green function techniques
and an explicit computation. The next result, Theorem \ref{thm: R
< sigma at high k}, is in the high-frequency regime. It can be
easily deduced from earlier results, e.g.\ \cite{Van0,Leis},
relying on the method of stationary phase.
\begin{theorem}\label{thm: R < sigma at high k}
Assume that the smooth, compact body $\Omega$ is strictly convex. There is $k_0>0$ such that for all $k>k_0$
\begin{equation}\label{anisH}
\sigma_{T}<\sigma
\end{equation}
\end{theorem}

\begin{remark}
The relation between the scattering problem presented in Section
\ref{sec: intro} and the boundary value problem as presented above, is given as
\begin{equation}\label{def: u}
\Psi (\mathbf{r})=\e^{\i k z}+u(\mathbf{r})
\end{equation}
\end{remark}

\begin{remark}
The condition that $\Om$ is strictly convex, assures that $f_{cl}$
exists. For example, if $\Om$ is a cylinder with axis
$\mathbf{e}$, then $f_{cl}$ doesnot exist, nevertheless
$R_{\mathrm{cl}},\sigma_{\mathrm{cl}}$ can still be defined by
\eqref{def: straight}, but now
$R_{\mathrm{cl}}=2\sigma_{\mathrm{cl}}$, which is the highest possible value for $R_{\mathrm{cl}}$.
\end{remark}
%\begin{equation}\label{anisH}
%R_{\mathrm{cl}}=\sigma_{\mathrm{cl}}
%\end{equation}

%\begin{remark}
%The assumption that $\Om$ is $C^{\infty}$ is probably
%unnecessarily restrictive. For example, the high-frequency
%estimates in \cite{majda} require $C^2$, the uniqueness of the
%solution of the boundary value problem in \cite{Ramm} requires
%even less than $C^2$. We make this assumption since we rely on
%\end{remark}
%
%
%
%   PROOFS
%
%
%
%
%
%

\section{Proofs} \label{sec: proofs}

\subsection{Proof of Theorem \ref{thm: perturbative low freq} }

For bodies $\Om$ with smooth boundary, one applies standard Green
function techniques, see e.g. \cite{tych}, to rewrite $u$, as
defined in Section \ref{sec: results}, in the form
\begin{equation}\label{def: u as integral}
u(\mathbf{r})=\int_{\partial \Omega} \d \sigma(\mathbf{p}) \,
\mu(\mathbf{p}) \frac{\e^{\i
k|\mathbf{p}-\mathbf{r}|}}{|\mathbf{p}-\mathbf{r}|} , \quad
\mathbf{r} \in \mathbb R^3 \setminus \Omega,
\end{equation}
where $\mu$ is given as the jump in normal derivative of $u$ on
$\partial \Om$,
\begin{equation}\label{eq: jump derivative}
 \mu (\mathbf{p})  =-\lim_{\footnotesize{\left. \begin{array}{c} \mathbf{r} \ra \mathbf{p} \\ \mathbf{r} \in \bbR^3 \setminus \Om \end{array}\right.}}\frac{\partial u}{\partial \mathbf{n}}(\mathbf{p})  + \lim_{\footnotesize{\left. \begin{array}{c} \mathbf{r} \ra \mathbf{p} \\ \mathbf{r} \in  \Om \end{array}\right.}}\frac{\partial u}{\partial
 \mathbf{n}}(\mathbf{p})
\end{equation}
where $\mathbf{n}$ is the outward normal at $\mathbf{p} \in
\partial \Om$ and $\frac{\partial }{\partial \mathbf{n}}$ stands for $\mathbf{n} \cdot \nabla$.  The connection between the \emph{scattering amplitude} $f$ and $\mu$ is given by
\beq \label{connection f and mu} f( \mathbf{q})= \int \d \si
(\mathbf{p})  \, \e^{-\i k \mathbf{p}\cdot \mathbf{q}
}\mu(\mathbf{q}) \eeq
Our strategy will be to expand the functions $u(\mathbf{r})$ and
$f(\mathbf{q})$ in powers of the wave number $k$ and to
investigate the behavior of $|f(\mathbf{q})|^2$ up to order $k^3$.
The formal expansions in powers of $k$ are justified by results in
\cite{Ramm,Van0} (in particular paragr. 2 Ch. 9 in \cite{Van0})
assure that the expansions  (\ref{decm}, \ref{decf}) are
convergent for all $k$.

We expand the function $\mu$ and $f$ up to $\caO(k^2)$,
\begin{equation}\label{decm}
\mu (\mathbf{p}) =\mu_0 (\mathbf{p}) +ik\mu_1 (\mathbf{p}) +(ik)^2
\mu_2 (\mathbf{p}) +\caO(k^3), \quad \mathbf{p} \in
\partial \Omega,
\end{equation}
\begin{equation}\label{decf}
f( \mathbf{q})=
%\frac{1}{4\pi}
f_0( \mathbf{q})+ik f_1( \mathbf{q})+(ik)^2 f_2( \mathbf{q})
+\caO(k^3) , \quad \mathbf{q} \in S^2,
\end{equation}
By using the boundary condition $u|_{\partial \Om}=- \e^{\i kz}$
and \eqref{connection f and mu}, we have
\begin{equation} \label{expansion: mu 0}
\int_{\partial \Omega} \d \sigma (\mathbf{p})\, \frac{\mu_0
(\mathbf{p})}{|\mathbf{p}-\mathbf{r}|} =-1, \quad \mathbf{r} \in
\partial \Omega,
\end{equation}
\begin{equation} \label{expansion: mu 1}
\int_{\partial \Omega}  \d \sigma (\mathbf{p}) \,\frac{\mu_1
(\mathbf{p})}{|p-r|}\d \si (\mathbf{p}) +\int_{\partial \Omega} \d
\sigma (\mathbf{p}) \, \mu_0 (\mathbf{p}) =-z, \quad \mathbf{r}
\in
\partial \Omega,
\end{equation}
\begin{equation} \label{expansion: mu 2}
\int_{\partial \Omega} \d \si (\mathbf{p})\, \frac{\mu_2
(\mathbf{p}) }{|p-r|} +\int_{\partial \Omega}  \d \si (\mathbf{p})
\, \mu_1 (\mathbf{p}) +\frac{1}{2}\int_{\partial \Omega} \d \si
(\mathbf{p}) \,  \mu_0 (\mathbf{p}) |\mathbf{p}-\mathbf{r}|
=-\frac{z^2}{2}, \quad \mathbf{r} \in
\partial \Omega,
\end{equation}
We evaluate the \emph{scattering amplitude} $f$,
\begin{equation} \label{expansion: f 0}
f_0( \mathbf{q})=\int_{\partial \Omega}  \d \si (\mathbf{p}) \,
\mu_0(\mathbf{p})
\end{equation}
\begin{equation} \label{expansion: f 1}
f_1( \mathbf{q})=-\int_{\partial \Omega}  \d \si (\mathbf{p}) \,
\mu_0(\mathbf{p}) (\mathbf{p}\cdot \mathbf{q}) +\int_{\partial
\Omega} \d \si (\mathbf{p}) \, \mu_1 (\mathbf{p})
\end{equation}
\begin{equation} \label{expansion: f 2}
f_2( \mathbf{q}) = \int _{\partial \Omega} \d \si (\mathbf{p}) \,
\mu_2 (\mathbf{p})  - \int_{\partial \Omega}  \d \si (\mathbf{p})
\, \mu_1 (\mathbf{p}\cdot \mathbf{q})  +\frac{1}{2} \int _{\partial \Omega}
\d \si (\mathbf{p}) \, \mu_0(\mathbf{p}) (\mathbf{q} \cdot
\mathbf{p} )^2 ,
\end{equation}
Let $\refl $ denote the inversion $z \ra -z$, acting on subsets of
$\bbR^3$. In particular
\begin{equation}
(\mathbf{x},z) \in \refl \Omega \Leftrightarrow (\mathbf{x},-z)
\in  \Omega
\end{equation}
We split a function $g$ on $\partial \Omega$ into `symmetric' and
`antisymmetric' parts as follows \beq g^s(\mathbf{p})
=\frac{1}{2}[ g (\mathbf{p},\Omega)+ g ( \refl \mathbf{p},\refl
\Omega) ] \qquad g^a(\mathbf{p})    =\frac{1}{2} [g
(\mathbf{p},\Omega)- g (\refl \mathbf{p},\refl \Omega)], \qquad
\mathbf{p} \in
\partial \Omega \eeq
and similarly for functions $h$ on $S^2$: \beq h^s ( \mathbf{q})
=\frac{1}{2}[ h ( \mathbf{q})+ h ( \refl \mathbf{q}) ] \qquad h^a
( \mathbf{q}) =\frac{1}{2} [h ( \mathbf{q})- h (\refl
\mathbf{q})], \qquad \mathbf{q} \in S^2 \eeq
With these definitions, we can immediately state:
\begin{equation}\label{easy conseq: 1}
\mu_0 =\mu^s_0   \qquad C :=  - f_0 \textrm{ is constant}
\end{equation}
\begin{equation} \label{easy conseq: 2}
   \int_{\partial \Omega} \d \si (\mathbf{p}) \, \frac{\mu^a_1(\mathbf{p}) }{|\mathbf{p}-\mathbf{r}|}
=-z
\end{equation}
\begin{equation}   \int_{\partial \Omega} \d \si (\mathbf{p})
\,\frac{\mu^s_1(\mathbf{p}) }{|\mathbf{p}-\mathbf{r}|}=C
\Rightarrow \mu^s_1= -C \mu_0
\end{equation}
\begin{equation} \label{easy conseq: 3}
\int_{\partial \Omega} \d \si (\mathbf{p}) \,
\frac{\mu^a_2(\mathbf{p}) }{|\mathbf{p}-\mathbf{r}|}
=-\int_{\partial \Omega} \d \si (\mathbf{p}) \,
\mu^a_1(\mathbf{p}) , \quad \mathbf{r} \in
\partial \Omega
\end{equation}
\begin{eqnarray} \label{easy conseq: 4}
f^a_2 (\mathbf{q})&=& -\cos \theta \int_{\partial \Omega} \d \si (\mathbf{p}) \,
\mu^s_1(\mathbf{p}) z(\mathbf{p})=- \cos \theta \int_{\partial \Omega} \d \si (\mathbf{p}) \,
\mu^a_1(\mathbf{p}) z(\mathbf{p}) \\
&=& \cos \theta C  \int_{\partial \Omega} \d \si (\mathbf{p}) \,
\mu^s_0(\mathbf{p}) z(\mathbf{p})- \cos \theta \int_{\partial \Omega} \d \si (\mathbf{p}) \,
\mu^a_1(\mathbf{p}) z(\mathbf{p})
\end{eqnarray}

\noindent Let $u,v$ be harmonic functions on $\bbR^3 \setminus \Omega$,
satisfying the boundary conditions \beq v |_{\partial \Om}=-z
 \qquad u |_{\partial \Om}=-1  \eeq
and apply Green's theorem
\begin{equation}\label{applied green}
\int_{R} \d \mathbf{x}\, (u \triangle v-v \triangle u)  =
\int_{\partial R} \d \si (\mathbf{p}) \,    (u \frac{\partial
v}{\partial \mathbf{n}}-v \frac{\partial u}{\partial \mathbf{n}})
\end{equation}
with $R$ being a smooth region in $\bbR^3 \setminus \Omega$,
infinitesimally close to $
\partial \Omega$ and extending far enough at infinity.
The left-hand side of \eqref{applied green} vanishes, the right
hand side gives
\begin{equation} \label{def: K}
K (\Omega) := \int_{\partial \Omega} \d \si (\mathbf{p}) \,
\mu^a_1(\mathbf{p})    = \int_{\partial \Omega} \d \si
(\mathbf{p}) \, z(\mathbf{p})   \mu_0(\mathbf{p})
\end{equation}
For $\mathbf{q}, \mathbf{p} \in \bbR^3$, we write $z(\mathbf{p})
,z( \mathbf{q})$ for their projections on the $z$-axis and
$\mathbf{p}^{\bot},\mathbf{q}^{\bot}$ for their projections on the
$\mathbf{e}^{\bot}$-plane. Recall also that $\cos \theta
=\mathbf{e} \cdot \mathbf{q}$. It follows that $\mathbf{p} \cdot
\mathbf{q}= z(\mathbf{p}) \cos \theta  + \mathbf{q}^{\bot} \cdot
\mathbf{p}^{\bot}$. Inserting \eqref{def: K} in \eqref{expansion:
f 1} yields
\begin{eqnarray} \label{general expression: f 1}
f_1( \mathbf{q}) &=&(1-\cos \theta) K  +  \int_{\partial \Omega}\d \si (\mathbf{p}) \, \mu^s_1(\mathbf{p}) -  \int_{\partial \Omega} \d \si (\mathbf{p}) \, (\mathbf{q}^{\bot} \cdot \mathbf{p}^{\bot}) \mu_0(\mathbf{p})   \\
&=& (1-\cos \theta) K  +  C^2- \int_{\partial \Omega} \d \si
(\mathbf{p}) \, (\mathbf{q}^{\bot} \cdot \mathbf{p}^{\bot})
\mu_0(\mathbf{p})
\end{eqnarray}
Remark that by \eqref{easy conseq: 3} and \eqref{def: K}
\begin{equation} \label{remark}
 \int_{\partial \Omega} \d \si (\mathbf{p}) \,    \mu^a_2(\mathbf{p})   = -C K(\Omega)
\end{equation}
%and that
%\begin{eqnarray} \label{remark}
% f_2(q,\Omega)-f_2(q,\refl \Omega)&=&-2C K (\Omega)-2 \int_{\partial \Omega}\d \si (\mathbf{p}) \,    \mu^s_1(\mathbf{p})    (q,p)\\ &=&-2C K (\Omega)+2C\cos \theta K -2\int_{\partial \Omega}\d \si (\mathbf{p}) \,    \mu_0(\mathbf{p})    (\mathbf{q}^{\bot} \cdot \mathbf{p}^{\bot})
%\end{eqnarray}
We expand the scattering amplitude up to $\caO(k^3)$;
\begin{equation} \label{eqn: |f|^2 function}
|f( \mathbf{q})|^2=f_0^2( \mathbf{q})   -2 k^2 f_0 ( \mathbf{q})
f_2( \mathbf{q})+ k^2 f^2_1( \mathbf{q})  + \caO(k^4)
\end{equation}
and we use the above estimates to obtain
\begin{eqnarray} \label{eqn: difference R - sigma}
&& \si-\sigma_{T}= \int_{S^2} \d \mathbf{q} \,  \cos \theta |f( \mathbf{q})|^2 \\
&=& 2 k^2 \int_{S^2}\d \mathbf{q} \,  \cos \theta \left( - f^s_0 f^a_2 + f^s_1 f^a_1    \right)  \nonumber \\
&=& 2 k^2 \int_{S^2} \d \mathbf{q} \,  \cos \theta  \Bigg\{  C
 \left(  \cos \theta C  \int_{\partial \Omega} \d \si (\mathbf{p}) \,
\mu^s_0(\mathbf{p}) z(p)- \cos \theta \int_{\partial \Omega} \d
\si (\mathbf{p}) \,
\mu^a_1(\mathbf{p}) z(p)    \right)   \nonumber   \\
&-& \cos \theta K  (K+C^2 -\int_{\partial
\Omega}  \d \si (\mathbf{p}) \,  (\mathbf{q}^{\bot} \cdot \mathbf{p}^{\bot}) \mu_0(\mathbf{p})       ) \Bigg\}  \nonumber \\
&=& 2k^2 \int_{S^2} \d \mathbf{q} \, \cos^2 \theta   \left(  C^2 K
-C \int_{\partial \Omega} \d \si (\mathbf{p}) \,
\mu^a_1(\mathbf{p}) z(p)-K^2 -K C^2 +K \int_{\partial
\Omega}  \d \si (\mathbf{p}) \,  (\mathbf{q}^{\bot} \cdot \mathbf{p}^{\bot}) \mu_0(\mathbf{p})     \right) \nonumber \\
&=& - \frac{4 \pi }{3}k^2   \left(   C \int_{\partial \Omega} \d
\si (\mathbf{p}) \, \mu^a_1(\mathbf{p}) z(p) +K^2     \right)
\nonumber
\end{eqnarray}
To obtain the last equality we used that   \beq \label{eq: extra
assymetry}  \int_{\partial \Omega}  \d \si (\mathbf{p}) \,
\mu_0(\mathbf{p}) \int_{S^2} \d \mathbf{q} \, \cos^2 \theta\,
(\mathbf{q}^{\bot} \cdot \mathbf{p}^{\bot}) \eeq vanishes since
the second integrand is antisymmetric with respect to the
transformation $(z(\mathbf{q}),\mathbf{q}^{\bot}) \rightarrow
(z(\mathbf{q}),-\mathbf{q}^{\bot})$. The rest of the proof will
consist in showing that \beq \label{rest of proof}- \frac{4 \pi
}{3}k^2 \left( C \int_{\partial \Omega} \d \si (\mathbf{p}) \,
\mu^a_1(\mathbf{p}) z(p) +K^2 \right) \geq \frac{4 \pi }{3}k^2 CV
 \eeq
 which immediately yields Theorem \ref{thm: perturbative low
 freq}.

%
%Using again \eqref{eq: extra assymetry} and $K(\Omega)=-K(\refl
%\Omega)$, we get
%
%\begin{eqnarray} \label{eqn: |f|^2  - R|f|^2}
%&& |f(q,\Omega)|^2-|f(q,\refl \Omega)|^2\\
%&=&- k^2 2 f_0(q,\Omega)(f_2(q,\Omega)-f_2(q,\refl \Omega)) + k^2(f_1(q,\Omega)+f_1(q,\refl \Omega))(f_1(q,\Omega)-f_1(q,\refl \Omega)) + \caO(k^4)  \nonumber \\
%&=& k^2 2 C \big(-2C K (\Omega)+2C\cos \theta K -2\int_{\partial \Omega}\d \si (\mathbf{p}) \,    \mu_0(\mathbf{p})    (\mathbf{q}^{\bot} \cdot \mathbf{p}^{\bot}) \big)\\
%&+& 4 \big(C^2 -\int_{\partial \Omega}(\mathbf{q}^{\bot} \cdot \mathbf{p}^{\bot})
%\mu_0(\mathbf{p})   \big) \big((1-\cos \theta)2K(\Omega) \big)+
%\caO(k^4) \nonumber
%\end{eqnarray}
%leading to
%\begin{equation} \label{eqn: sigma  - R sigma}
%\si (\Omega)- \si (\refl \Omega)= 16\pi k^2 C^2 K
%\end{equation}
%R(\Omega)- R(\refl \Omega)&= &    -5k^2 C^2  K    \frac{4 \pi}{3}

 Let
\begin{equation}
v(\mathbf{r}) = \int_{\partial \Om}\d \si (\mathbf{p}) \,
\frac{\mu^a_1(\mathbf{p}) }{|\mathbf{p}-\mathbf{r}|}, \qquad
v_{int}=v |_{\Om}, \quad  v_{ext}=v |_{\bbR^3 \setminus \Om}
\end{equation}
Both $v_{ext}$ and $v_{int}$ are harmonic functions which can be
continuously extended to $\partial \Omega$. We know that
$v|_{\partial \Om}=-z$ and hence necessarily $v_{int}=-z$. By
Green function techniques (compare with \eqref{eq: jump
derivative}), we have
\begin{equation}\label{eq: jump derivative}
 \mu^a_1(\mathbf{p})    =-\lim_{\footnotesize{\left. \begin{array}{c} \mathbf{r} \ra \mathbf{p} \\ \mathbf{r} \in \bbR^3 \setminus \Om \end{array}\right.}}\frac{\partial v_{ext}}{\partial \mathbf{n}}(\mathbf{p})    + \lim_{\footnotesize{\left. \begin{array}{c} \mathbf{r} \ra \mathbf{p} \\ \mathbf{r} \in  \Om \end{array}\right.}}\frac{\partial  v_{int}}{\partial
 \mathbf{n}}(\mathbf{p})
\end{equation}
Calculate
\begin{eqnarray}\label{aux2}
-\int_{\partial \Omega} \d \si(\mathbf{p}) \,
\mu^a_{1}(\mathbf{p})z (\mathbf{p})&=& \int_{\partial \Omega}\d
\sigma(\mathbf{p})\, v_{ext}(\mathbf{p})    \left ( \frac{\partial
v_{int}} {\partial \mathbf{n}}(\mathbf{p})-
\frac{\partial v_{ext}} {\partial \mathbf{n}}(\mathbf{p}) \right )    \\
&=&\int_{\partial \Omega}\d \sigma(\mathbf{p})\, z(\mathbf{p})
\frac{\partial z(\mathbf{p})} {\partial \mathbf{n}}
-\int_{\partial \Omega} \d \sigma( \mathbf{p})\,
v_{ext}(\mathbf{p}) \frac{\partial v_{ext}    }
{\partial \mathbf{n}}(\mathbf{p})  \\
&=& V +\int_{\mathbb R^3 \backslash  \Omega} \d \mathbf{r}\,|
\nabla v_{ext}(\mathbf{r})|^2   ,
\end{eqnarray}
where $V$ is the volume of $\Omega$. To get the last equality, we
applied the divergence theorem.\\
%This ends the proof of item
%(1) of Theorem \ref{thm: perturbative low freq}.\\
%
Put
\[
M (\Omega)= -\int_{\partial \Omega} \d \si (\mathbf{p}) \,
z(\mathbf{p}) \mu^a_1(\mathbf{p})     - V =\int_{\bbR^3 \setminus
\Om} \d \mathbf{r} \,|\nabla v_{ext}(\mathbf{r})  |^2
\]
Define also
\begin{equation}
u(\mathbf{r}) = \int_{\partial \Omega} \d \si (\mathbf{p}) \,
\frac{\mu_0(\mathbf{p}) }{|\mathbf{p}-\mathbf{r}|},  \qquad
u_{int}=u \big|_{\Om}, \quad u_{ext}=u \big|_{\bbR^3 \setminus
\Om}
\end{equation}
Reasoning as above, we have that $u_{int}=-1$, hence
\begin{eqnarray}
-C=\int_{\partial \Omega} \d \si(\mathbf{p}) \,
\mu_{1}(\mathbf{p})&=&- \int_{\partial \Omega}\d
\sigma(\mathbf{p})\, u_{ext}    \left ( \frac{\partial u_{int}}
{\partial \mathbf{n}}(\mathbf{p})  -
\frac{\partial u_{ext}} {\partial \mathbf{n}}(\mathbf{p})  \right )    \\
&=& -\int_{\mathbb R^3 \backslash  \Omega} \d \mathbf{r}\,| \nabla
u_{ext} (\mathbf{r})  |^2   ,
\end{eqnarray}
and
\begin{eqnarray}
K=\int_{\partial \Omega} \d \si(\mathbf{p}) \, z(\mathbf{p})
\mu_{1}(\mathbf{p})&=& \int_{\partial \Omega}\d
\sigma(\mathbf{p})\, v_{ext}(\mathbf{p})
\frac{\partial u_{ext}} {\partial \mathbf{n}} (\mathbf{p})   \\
&=& \int_{\bbR^3 \setminus \Om} \d \mathbf{r} \, (\nabla u_{ext}
(\mathbf{r}) \cdot \nabla v_{ext}(\mathbf{r}))     ,
\end{eqnarray}
%Reasoning as above, we get \beq  C= \int_{\bbR^3 \setminus \Om} \
%\d \mathbf{r} \, | \nabla u_{ext} |^2  \qquad   K= \int_{\bbR^3
%\setminus \Om} \d \mathbf{r} \, \nabla u_{ext} \cdot \nabla
%v_{ext}    \eeq

Since the functions $\nabla u_{ext}, \nabla v_{ext}$ are square
integrable, the Cauchy-Schwarz inequality yields  \beq
\label{Cauchy schwarz consequence} K^2 (\Omega) \leq M (\Omega)  C
(\Omega) \eeq which means that in \eqref{eqn: difference R -
sigma}, we can estimate \beq C(M+V)-K^2 \geq C(M+V)-CM = CV  \eeq
which proves the inequality \eqref{rest of proof} since the LHS of
\eqref{rest of proof} is $\frac{4 \pi }{3}k^2\left( C(M+V)-K^2
\right)$. This ends the proof.
\subsection{Proof of Theorem \ref{thm: R < sigma at high k}}

From techniques, based on the method of stationary phase, we know
(see \cite{majda}) that for strictly convex
bodies $\Om$,
\begin{equation}\label{sR}
|f( \mathbf{q})|^2=|f_{\mathrm{cl}}( \mathbf{q})|^2+O(1/k^2)
,\qquad  S^2 \ni \mathbf{q} \neq \mathbf{e}
\end{equation}
where the error estimate $O(1/k^2)$ is uniform in every compact
subset of $ S^2$ which does not contain $\mathbf{e}$. From
\cite{Leis}, we know that
\begin{equation}\label{sR2}
\lim_{k \rightarrow \infty} \int_{S^2} \d \mathbf{q} \, |f(
\mathbf{q})|^2 =2\sigma_{\mathrm{cl}}
\end{equation}
Combining \eqref{sR}, \eqref{sR2} and \eqref{def: classical R and
sigma}, we get, in the sense of distribution on $S^2$,
\begin{equation}\label{gen}
\lim_{k \uparrow \infty} |f|^2= |f_{\mathrm{cl}}|^2+ |\mathcal I |
\delta_{\mathbf{e}}
\end{equation}
where $\delta_{\mathbf{e}}$ is the Dirac delta distribution on
$S^2$, peaked at $\mathbf{e} \in S^2$. An immediate consequence is
\begin{equation}
\lim_{k \rightarrow \infty} \sigma_{T} = R_{\mathrm{cl}}.
\end{equation}
From the definition of $\sigma_{T}$ and $\si$ follows
\begin{equation}
\sigma_{T} \leq 2\si, \qquad R_{\mathrm{cl}} \leq 2
\si_{\mathrm{cl}}
\end{equation}
The second inequality is an equality only when the side of $\Om$,
exposed to the incoming flow, is perpendicular to $\mathbf{e}$.
Since we exclude this by assuming strict convexity, we get
\begin{equation}
R_{\mathrm{cl}}<2\si_{\mathrm{cl}}
\end{equation}
Let  $\eps=(2\si_{\mathrm{cl}}-R_{\mathrm{cl}})/2$. Using
\eqref{gen}, we find a $k_0 >0$ such that for
 $k>k_0$
\begin{equation}
\sigma_{T}<R_{\mathrm{cl}}+\eps, \qquad
\si>2\si_{\mathrm{cl}}-\eps,
\end{equation}
and hence
\begin{equation}
\sigma_{T}<R_{\mathrm{cl}}+\eps=2\sigma_{\mathrm{cl}}-\eps<\si
\end{equation}
which ends the proof.\\

\noindent \textbf{Acknowledgement} A significant part of this work
was done during a visit of E.L.L. to TUM (Technical University of
Munich) and RUB (Ruhr-University of Bochum) in 2005. E.L.L. thanks
both institutes and Frau Cristel Schr\"oder for warm hospitality.
E.L.L. was also supported by {\it Centre for Research on
Optimization and Control (CEOC)} from the ''Funda\c{c}\~{a}o para
a Ci\^{e}ncia e a Tecnologia" FCT, cofinanced by the European
Community Fund FEDER/POCTI.

%And
%also very thankful to frau Cristel Schr\"oder for her warm
%hospitality and care for my family.

\end{document}